# Theoretical study of ternary silver fluorides $AgMF_4$ (M = Co, Ni, Cu) formation at pressures up to 20 GPa

M.A. Domański[1]*, M. Derzsi[1,2] and W. Grochala[1]*

[1]*Centre of New Technologies, University of Warsaw, S. Banacha 2C, 02-097 Warsaw, Poland*
[2]*Advanced Technologies Research Institute, Faculty of Materials Science and Technology in Trnava, Slovak University of Technology in Bratislava, 917 24, Trnava, Slovakia*

*e-mail: m.domanski@cent.uw.edu.pl , w.grochala@cent.uw.edu.pl

***This work is dedicated to Prof. Gary Schrobilgen at his 75th birthday***

## Abstract

Only several compounds bearing Ag(II) cation and other transition metal cation have been known. Herein, we predict stability and crystal structures of hypothetical ternary silver(II) fluorides with copper, nickel and cobalt in 1:1 stoichiometry at pressure range from 0 GPa up to 20 GPa within the frame of Density Functional Theory. Calculations show that $AgCoF_4$ could be synthesized already at ambient conditions but this compound would host diamagnetic Ag(I) and high-spin Co(III). However, at increased pressure ternary fluorides of Ag(II) featuring Cu and Ni could be synthesized, in the pressure windows of 7-14 and 8-15 GPa, respectively. All title compounds would be semiconducting and magnetically ordered.

## 1. Introduction

Recent experimental and theoretical research on $AgF_2$ demonstrated that this material mimic very closely the key structural and electronic properties of the well-known lanthanum oxocuprate[1] (the prototype precursor of high-temperature oxocuprate superconductors). Pristine $AgF_2$ exhibits strong antiferromagnetic coupling *via* superexchange mechanism in two-dimensional sheets[2] and substantial mixing of ligand and metal states in the top of the valence band. Because of the striking similarities, not observed in any other known compound, it was suggested that the success of oxocuprates could possibly be repeated by related silver(II) fluorides if chemical doping to $AgF_2$ could be realized[1,3,4]. It is therefore highly desirable to examine various possibilities to modify its electronic and magnetic properties *via* doping.

Likewise the oxocuprates, where appropriate doping is required in order to exhibit superconductivity, also fluoroargentates could be doped to achieve this goal. Inserting the surplus holes or electrons is the necessary step towards emergence superconducting transition. However, in contrast to cuprates, the bulk $AgF_2$ has so far resisted all doping attempts. The hole doping of $AgF_2$ could be introduced by $F^-$ vacancies, but theoretical studies show that doping to this 001-type system *via* addition or depletion of its F contents does not lead to stable metallic phase[5]. Partial oxidation of Ag(II) is also difficult, due to immensely strong oxidative character of Ag(III). Therefore, our attention turned to the electron doping, which could be introduced in two ways. The first one could be provided by substitution at anionic sites ($F^- \rightarrow O^{2-}$), which will be described elsewhere. Here, we focused on the second possibility, where doping may be provided by electron injection *via* substitution at cationic site. Therefore, we considered diverse ternary fluoride stoichiometries which may host strongly coupled $AgF_2$ sheets. Among ternary fluorides $Ag^{II}M_xF_y$, many transition metal systems have been successfully prepared so far[3], *i.e.* for M = Au(III), Au(V), Nb(V), Ta(V), Ti(IV), Zr(IV), Hf(IV), Ru(V), Rh(IV), Ir(V), Pd(IV), Pt(IV), Pt(V), Mn(IV) and Cr(IV) ($AgMn^{IV}F_6$[6] and $AgCr^{IV}F_6$[7] structures not known). The great majority of these compounds contain closed-shell or low-spin M cations. Yet none of these fluorides host [$AgF_2$] sheets, as required for doping.

Herein, we theoretically search for systems bearing divalent cations which could possibly host [$AgF_2$] sheets using Density Functional Theory (DFT) in combination with evolutionary algorithms for crystal structures' prediction. We examine $AgMF_4$ systems with transition metals M = Cu, Ni and Co. These late 3d transition metals have been selected due to their relative resistance towards oxidation in the fluoride environment. The redox potentials for the $M^{3+}/M^{2+}$ redox pairs decrease in the Cu-Ni-Co sequence. $CoF_3$ and $NiF_3$ (actually $Ni^{II}Ni^{IV}F_4$[8]) are moderately stable, and they constitute strong fluorinating agents. $CuF_3$ is rather unstable[9] (even more than $AgF_3$, which structure was determined[10]), thus minimizing the possibility of the intrinsic redox reaction between Ag(II) and Cu(II). Aside from variations of the chemical composition, we also study the impact of external pressure on crystal and electronic structure and stability of $AgMF_4$ systems with respect to the mixture of binary fluorides, $AgF_2$ and $MF_2$ (M=Co, Ni, Cu).

## 2. Methods

This theoretical study is based on periodic electronic-structure calculations carried out with VASP 5.4.4 software using PAW method[11,12]. We used potentials set recommended by VASP with 520 eV plane-wave energy cut-off. Energy was calculated using collinearly spin-polarized DFT method using GGA functional (PBEsol, *i.e.* solid-revised Perdew, Burke and Ernzerhof correlation-exchange functional[13]). Geometry optimizations were done with k-spacing of 0.024 $Å^{-1}$ and conjugate-gradient algorithm relaxation with convergence criteria of $10^{-7}$ eV (electronic cycle) and $10^{-5}$ eV (ionic cycle). The tetrahedron method with Blöchl corrections was used to calculate electronic density of states (DOS). On-site electronic correlation was included with Hubbard ($U$) and exchange repulsion ($J$) terms using Dudarev's approach[14,15]. In this approach effective $U_{\text{eff}} = U - J$, is considered, thus we used $U_{\text{eff}}$ equal 4 eV (for Ag, Ni), 5 eV (Co) and 8 (Cu). These values were earlier used and validated in respective systems exhibiting +II oxidation state[2,16–18]. Such choice of $U_{eff}$ is additionally justified by successful confirmation of the experimentally observed pressure induced phase transitions of $AgF_2$[19], $CuF_2$[20], $NiF_2$[21] and $CoF_2$[16,21] up to 20 GPa (note in **SI**, see also **Table S1** and **Figure S1**). For some systems, geometry optimization was additionally performed using the CPU-consuming HSE06 hybrid functional[22] with a coarser k-spacing (0.048 $Å^{-1}$).

Quest for a lowest-energy structures was performed using particle-swarm evolutionary algorithm, the crystal structure prediction was done utilizing XtalOpt software[23] in combination with DFT calculations using VASP optimizer. The XtalOpt was utilized *ca*. 60 times in solid state chemistry or physics applications[24], however most often for binary systems. The cross-check XtalOpt runs were performed for the three title stoichiometries with Z=2 or 4 structural units in the unit cell, both at ambient pressure and at 10 GPa, and in each case over 500 distinct structures were generated (with many more duplicates). Manual screening of the Inorganic Crystal Structure Database (ICSD) was also done in search for the appropriate $ABC_4$ structures, used also as seeds structures in XtalOpt searches. Final optimization of the lowest-energy structures from XtalOpt search and models from ICSD was performed with the above described DFT+U settings. Using XtalOpt, a four-step optimization of each structure was performed increasing accuracy at each step, with last two using GGA+U approach. VESTA[25] software was used for visualization of structures. Band structure plots were made using PyProcar library[26] and AFLOW[27] program.

## 3. Results

In our quest for the lowest-energy crystal structure of hypothetical compounds of $AgCuF_4$, $AgNiF_4$ and $AgCoF_4$, we have initially tested various substitutions within known crystal structures of $AgF_2$, $CuF_2$, $NiF_2$ and $CoF_2$ from ICSD database. Secondly, XtalOpt evolutionary algorithm was used for the same purpose and for cross-checking the results. Moreover, we have tested various prototypical $M^{I}M^{III}F_4$ structures, which allow for electron transfer between metal sites; notably, we presumed that electron-hungry Ag(II) could undergo reduction, with the concomitant 1- or even 2e-oxidation of its neighbouring TM cations. The three scrutinized 3d elements, Co, Ni and Cu, differ in terms of redox properties, electron count, U-induced splitting of upper and lower Hubbard d-bands, and degree of tetragonal distortion of $MF_6$ octahedra. All above-mentioned cations also have significantly smaller cationic radii than Ag(II). The hypothetical mixed-cation fluorides are described below one by one, first in terms of structure and electronic properties at ambient pressure. Next, we focus on the pressure-induced stabilization of the proposed ternary phases and also their electronic properties.

### Crystal structures and stability of mixed-cation fluorides at ambient pressure

Due to similar chemistry of Ag(II) and Cu(II) as well as similar crystal structures of binary fluorides, our primary candidate for the $AgCuF_4$ structure was the distorted $PdF_2$ type structure typical of $AgF_2$[28]

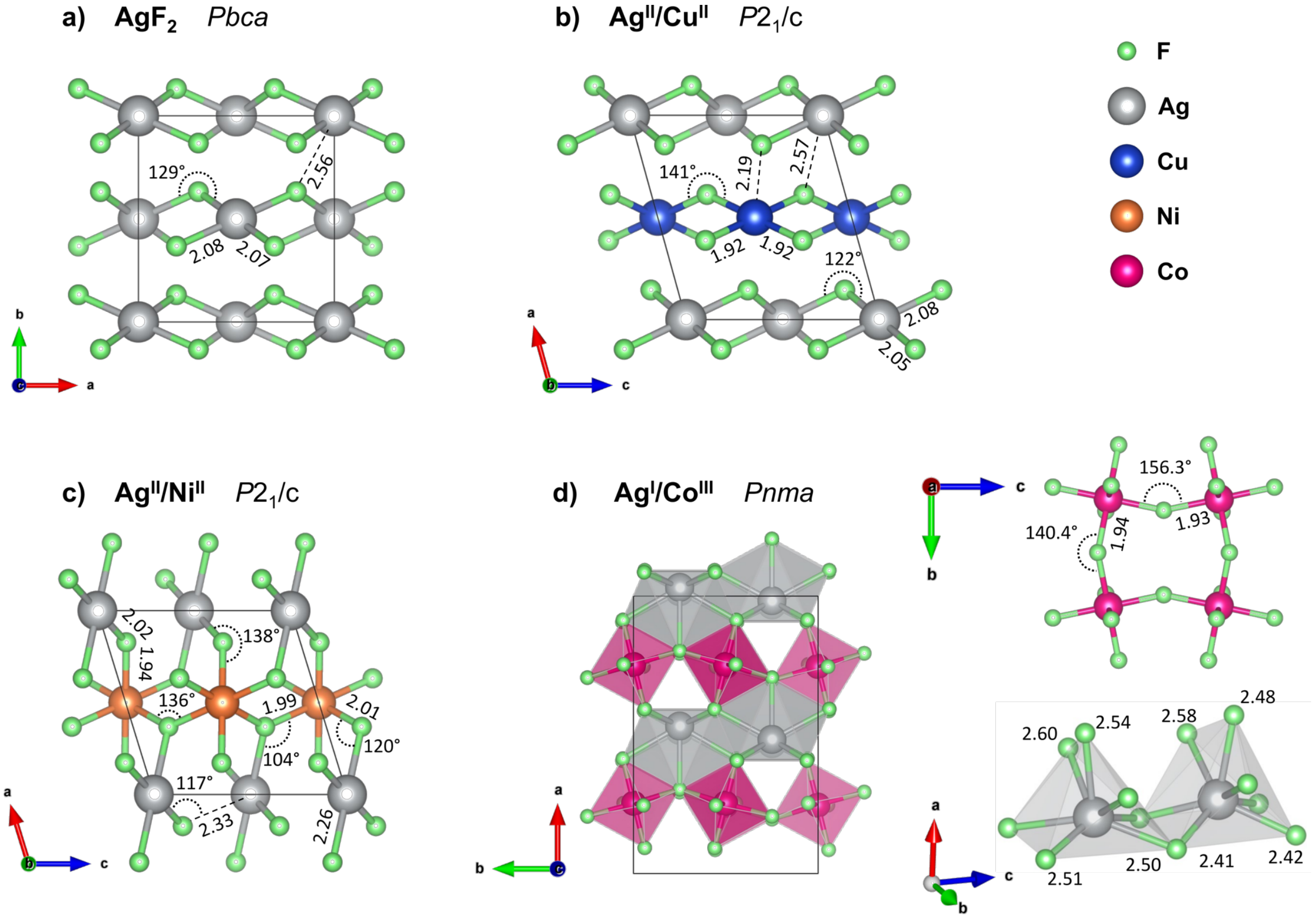

**Figure 1.** Crystal structures of the lowest-energy predicted $AgCoF_4$, $AgNiF_4$, $AgCuF_4$ and parent $AgF_2$ at ambient pressure (LP structures). The threshold for Ag-F bond drawing is 2.7 Å in $Ag^{I}Co^{III}F_4$ structure and 2.3 Å in all other cases. All bond lengths provided in angstroms.

(**Figure 1a**). Three types of substitutions at the metal site were considered, *i.e.* metal alternation in the directions perpendicular to the $\vec{a}$, $\vec{b}$ or $\vec{c}$ lattice vectors (*cf.* **SI**, **Figure S2**). Notably, these three diverse polymorphs were also found during XtalOpt search (see **SI, Figure S6**). We found that monoclinic distorted $AgF_2$ type structure with metal alternation in the *a* direction is the lowest-energy structure of $AgCuF_4$ (LP, **Figure 1b**) and this conclusion was confirmed using XtalOpt search; this polymorph preserves layered character of its binary constituents. Substantial tetragonal distortion of both Ag(II) and Cu(II) is predicted in agreement with the strong Jahn-Teller effect which is expected for $d^9$ species having coordination number (CN) of 6 . The tetragonal distorted octahedra usually constitute the (2 + 2) + 2 coordination of central atoms, which can be approximated to (4) + 2 for the most of cases. In such instance, we calculated the dimensionless Jahn–Teller distortion parameter, which equals the ratio $R = d_{ax}/d_{eq}$. In proposed $AgCuF_4$ the ratio of axial (2.570 Å) to equatorial (2.065 Å, averaged) bond lengths equals 1.24 for the Ag(II) site and 1.14 for the Cu(II) one (with axial bond lengths of 2.188 Å and equatorial ones of 1.919 Å). For comparison, the $R$ ratio equals to 1.23 and 1.18, for $AgF_2$ and $CuF_2$, respectively. The $AgCuF_4$ forms antiferromagnetic [$AgF_{4/2}$] and [$CuF_{4/2}$] separate layers inside crystal lattice (consequently, this polymorph is labelled as SL). The geometry within the puckered layers is reminiscent of those found in binary fluorides; the [$AgF_2$] layers are more buckled than in pure $AgF_2$ (the Ag-F-Ag angle is 122.0° *vs.* 129.1° in bulk $AgF_2$), while [$CuF_2$] layers are flattened (the corresponding angle is 140.8° *vs.* 131.8° in bulk $CuF_2$). This feature obviously stems from difference of the ionic radii between Ag(II) and Cu(II). In order to form a segregated layer structure without much strain the [$AgF_2$] sheets must buckle and the [$CuF_2$] sheets must simultaneously flatten; this accommodation is associated by bending of the crystallographic beta angle from 90° (for bulk $AgF_2$) to 105.5°. Interestingly, several higher energy crystal structures either contain Ag-F-Cu bridges and they preserve layered character (as it is the case for the polymorphs with metal substitutions $\perp \vec{a}$ and $\perp \vec{c}$), or they exhibit genuine 3D connectivity (**SI**, **Figure S2**). Importantly, the most stable SL polymorph is computed to be +6.7 kJ/mol uphill in energy with respect to substrates:

$$AgF_2 + CuF_2 \rightarrow AgCuF_4 \qquad \text{(Eq.1)}$$

Thus, it should not dissociate to the binaries at p → 0 GPa, T → 0 K conditions. If, however, it could be achieved in experiment, it would be metastable with respect to phase separation.

Predictions of the crystal structure of $AgNiF_4$ – in contrast to those featuring Cu(II) – were influenced by the fact that Ni(II) reveals a known tendency to be oxidized to Ni(III) or even Ni(IV) (though higher fluorides of nickel tend to release $F_2$).[29] However, during our quest only very few such structures with intervalence charge transfer were found, having considerably higher enthalpy (*i.e.* at least +0.620 eV, *cf.* **SI**, **Figure S6**). Similarly, as in the case of Cu(II), the final lowest-energy structure can be derived from parent $AgF_2$ type structure, however it is not layered. The electronic ground state of this ternary fluoride is antiferromagnetic. In contrast to the parent $NiF_2$ (tetragonal rutile structure) and $AgF_2$ the lowest-energy $AgNiF_4$ has monoclinic structure with beta angle 107.1°. In the mixed cation fluoride one can distinguish the Ag(II) site forming $AgF_6$ octahedron with shorter axial bond (2.025 Å) and two longer Ag-F bonds (2.260 and 2.332 Å, **Figure 1c**), which is octahedrally compressed with $R$ ratio equal *ca.* 0.88. The tetragonal compression of octahedra is well known from the fluorine chemistry of Ag(II).[30] On the other hand, the $NiF_6$ octahedra are nearly undistorted with axial bonds lengths of 1.941 Å and equatorial ones of 1.995 and 2.008 Å. In the LP structure Ag(II) cations act as connectors between $[NiF_{4/2}F_{2/1}]^{2-}$ layers present within the ***bc*** plane. However, the Ag-F bonds involved in formation of the Ag-F-Ni bridges are shorter than the Ag-F bonds forming Ag-F-Ag' bridges (**Table 1**). This LP structure was labelled 3D due to its spatial connectivity and its lowest-energy among all structures found at 0

GPa, but it still has energy +8.6 kJ/mol with respect to the binary substrates $AgF_2$ + $NiF_2$. In conclusion, if synthesized, $AgNiF_4$ could be metastable at ambient pressure.

In the case of $AgCoF_4$, the difference between standard reduction potentials for the $Co^{3+/2+}$ and $Ag^{2+/1+}$ redox pairs is considerable (1.82 V vs. 1.98 V[31], respectively), so a facile reaction:

$$AgF_2 + CoF_2 \rightarrow AgF + CoF_3 \qquad \text{(Eq.2)}$$

could be expected. Calculations show that this reaction would not be favoured though, on both DFT+U level of theory (+25.8 kJ/mol) or hybrid density functional HSE06 (+12.2 kJ/mol, see HSE06 results for all the binary and ternary compounds in **SI**, **Table S2**). Nevertheless, it appears that the lowest-energy obtained mixed-cation structure corresponds in fact to the $Ag^{I}Co^{III}F_4$ formulation; this is because enthalpy for the Lewis acid-Lewis base reaction:

$$AgF + CoF_3 \rightarrow AgCoF_4 \qquad \text{(Eq.3)}$$

is negative, some –30.1 kJ/mol at DFT level (–37.1 kJ/mol with hybrid functional). The spontaneous redox reaction (which manifests itself by disappearance of the spin on Ag site, and modification of magnetic moment on the Co site) is seen even for the models derived from parent $AgF_2$ type structure. The lowest-energy polymorph found was $KFeF_4$-LT[32] type structure (Z=8, **Figure 1d**). DFT+U calculations yield energy of this structure to be –4.2 kJ/mol (–24.8 at HSE06 level) with respect to the binary divalent metal fluorides:

$$AgF_2 + CoF_2 \rightarrow AgCoF_4 \qquad \text{(Eq.4)}$$

Further structure quests with XtalOpt resulted in many similar, layered $Ag^{I}Co^{III}F_4$ structures, including $KFeF_4$-HT polymorph (Z=4, for other see **SI**, **Figure S6**). The $KFeF_4$-LT type crystal structure comprises the puckered antiferromagnetic anionic sheets of $[CoF_{4/2}F_{2/1}]^-$ stoichiometry which feature high-spin Co(III) cations, and diamagnetic Ag(I) acting as counterion. While in parent $CoF_3$ the octahedra are almost perfectly symmetric (with all bond lengths of 1.886 Å, the F-Co-F angles are distorted ±1.5° from 90°), in $AgCoF_4$ the $CoF_6$ octahedra are markedly compressed with axial bond length 1.824 Å and two equatorial ones of 1.93–1.95 Å (ratio of *ca.* 0.94). The axial Co-F bonds are oriented towards Ag(I) layer, while equatorial bonds form a layer parallel to ***bc*** plane with angles 140.4° and 156.3° on Co-F-Co bridges; for comparison, the corresponding angles found in binary $CoF_3$ are 149.3°. Considering coordination sphere of Ag(I), pristine AgF crystalizes in rock-salt structure with perfect octahedra having 2.426 Å Ag-F bond lengths. In the case of $AgCoF_4$ however, two types of Ag(I) can be distinguished, but each Ag(I) has CN=7 within 2.65 Å radius which form bicapped tetragonal pyramid. One Ag(I) site exhibits 2.41-2.42 Å bond lengths to fluoride anions within tetragonal base, 2.482 Å in the apex and 2.576 Å for the two capping fluorides; the other Ag(I) site features 2.50-2.51 Å, 2.537 Å and 2.602 Å bond lengths, respectively. Two more distant fluoride ions are found at 2.767 Å.

## Crystal structures of high-pressure polymorphs at 10 GPa

Pressure may affect to a great extent both the crystal structures and stability of chemical compounds.[33] Here we have examined the impact of a rather moderate pressure of up to 20 GPa on formation of the title $AgMF_4$ compounds.

Pressure increase from 0 to 10 GPa in $AgCuF_4$ system turns out to promote the formation of the HP1 (3D) structure exhibiting spatial connectivity (**Figure 2b)** rather than the layers structure obtained at 0 GPa **Figure 1b**). The HP1 (3D) polytype is based on $AgF_2$ type structure with the shortest Ag-F bonds involved in the formation of strong Ag-F-Cu bridges (**Figure 2b**). This polymorph has already been identified as a low-volume polymorph at ambient pressure in our XtalOpt quest. It is enthalpically preferred over LP (SL) polymorph starting from about 6 GPa (**Figure 3a**). The HP1 form of $AgCuF_4$ has ferrimagnetic character with small uncompensated spin (*ca*. 0.02 $\mu_B$) due to opposite spin on Ag(II) and Cu(II) cations with slightly differing magnetic moments on each type of TM cation. Like in the low-pressure polymorph, here the Cu and Ag octahedra are orthorhombically distorted and tetragonally elongated, respectively. In the case of Ag-F bonds, the axial contacts are equal to 2.382 Å and the equatorial ones are 2.030 Å and 2.099 Å (with average ratio, $R$, of 1.15) while Cu-F bonds have three different lengths of 2.148 Å, 1.848 Å and 1.905 Å, respectively. As mentioned, dominant structural motifs in this crystal structure are the Ag-F-Cu bridges along directions [101] and [110] forming infinite antiferromagnetic chains. Chains along [110] feature 2.030 Å Ag-F and 1.848 Å Cu-F bond lengths with 133.6° angle while those along [101] are built from 2.099 Å Ag-F and 1.905 Å Cu-F bonds at 120.4° angle.

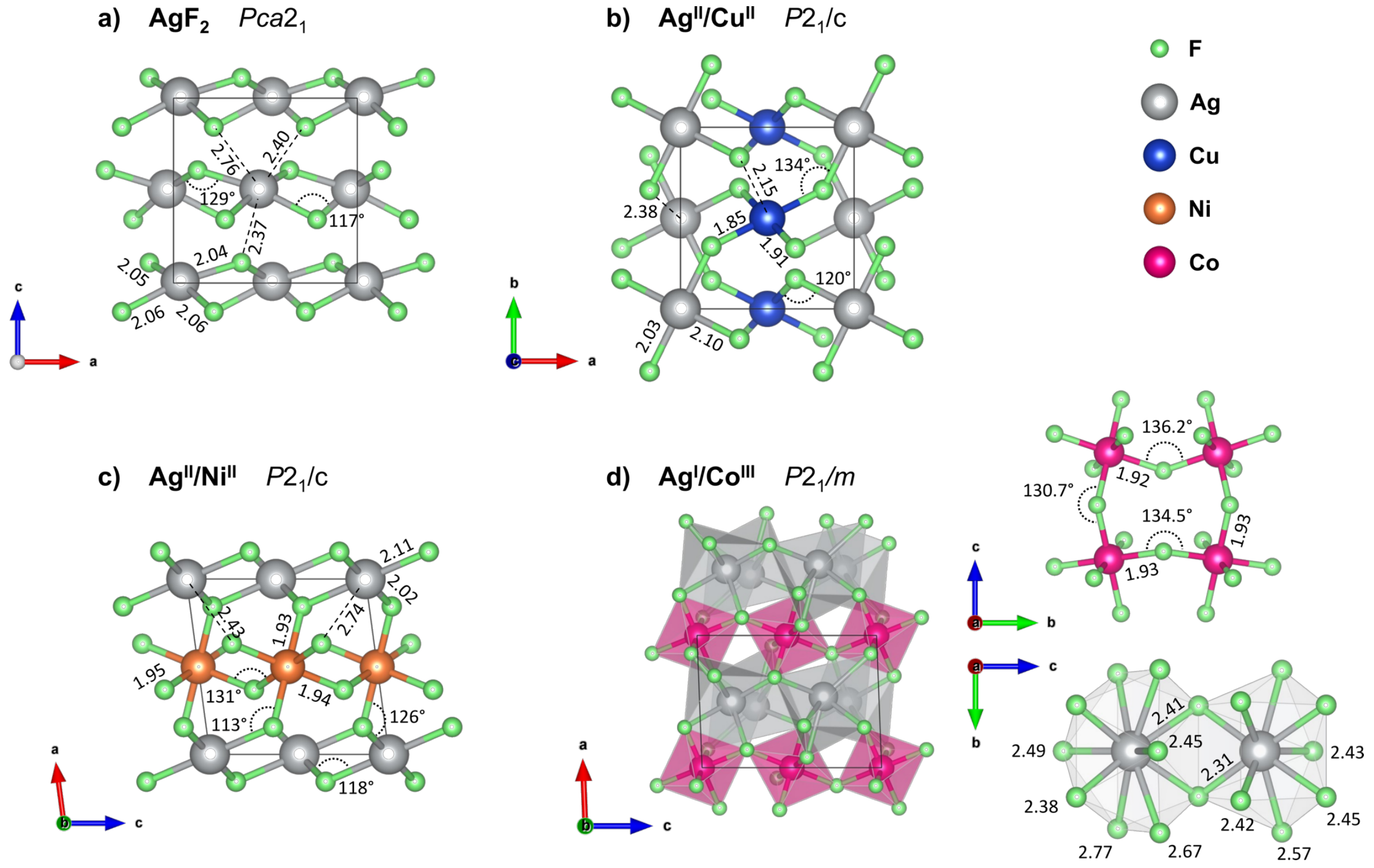

**Figure 2.** Crystal structures of the lowest-enthalpy predicted $AgCoF_4$, $AgNiF_4$, $AgCuF_4$ and parent $AgF_2$ at 10 GPa.

Also in $AgNiF_4$ system we predict change in relative stability of diverse polymorphs. In contrast to $AgCuF_4$, here the layered structure (HP1, **Figure 2c**) gains stability under pressure with respect to the 3D polymorph. The HP1 (SL) polymorph can be viewed as consisting of separate antiferromagnetic layers of $AgF_2$ and $NiF_2$ parallel to ***bc*** plane. In this structure, both Ag-F bonds within $AgF_2$ layer are considerably shorter (2.021 Å and 2.112 Å) than the axial Ag-F bond (2.433 Å, corresponding to an average $R$ factor of 1.18) which connects $AgF_2$ and $NiF_2$ layers. Also, the Ag-F-Ag angle within antiferromagnetic $AgF_2$ layer is rather far from straight (117.7°), and comparable with those computed for the HP1 polymorph of $AgF_2$ of 116.9° and 129.1° (**Figure 2a**). Simultaneously, the shortest Ni-F bonds at 1.927 Å connect Ni(II) with $AgF_2$ layer indicating a possibility of a strong magnetic interaction between both cations; the Ag-F-Ni angle equals 125.8°. Indeed, these magnetic centres form antiferromagnetic chains along [1 -1 0] direction. The HP1 (SL) structure of $AgNiF_4$ was found to be preferred over the LP polytype starting from 3 GPa and has negative enthalpy as compared to the binary fluorides from the pressure of 7 GPa (**Figure 3b**).

High pressure polymorph of $AgCoF_4$ found at 10 GPa preserves the layered-type structure found for the LP polymorph. The distorted monoclinic $KMnF_4$-related structure found using XtalOpt (HP1, **Figure 2d**) is preferred over the $KFeF_4$-LT type structure starting from the pressure of ~2 GPa (**Figure 3c**). This polymorph resembles LP polymorph in having antiferromagnetic puckered sheets of $[CoF_{4/2}F_{2/1}]^{-}$ stoichiometry with high-spin Co(III) cations and diamagnetic Ag(I) in between. Antiferromagnetic sheets are built from $CoF_6$ octahedra exhibiting tetragonal compression *i.e.* with short axial (1.78 Å) and longer equatorial bond lengths (1.92–1.93 Å) with $R$ factor of about 0.92 at 10 GPa. These equatorial bonds form a layer parallel to ***bc*** plane with Co-F-Co angles equal to 130.7°, 134.5° and 136.2°. The second constituent of $Ag^{I}Co^{III}F_4$ at 10 GPa, the Ag(I) cation, is found in two distinct crystallographic sites which differ in their coordination number. Here, CN is either 9 or 10, with bond lengths varying from 2.31 up to 2.77 Å (**Figure 2d**). At the same pressure, parent AgF transforms into CsCl type structure with cubic coordination sphere (8 ligands) and Ag-F bond length of 2.47 Å. Volume decrease of $Ag^{I}Co^{III}F_4$ with respect to binary fluorides is mainly due to enhancement of $[CoF_{4/2}F_{2/1}]^{-}$ buckling and increase of Ag(I) coordination number.

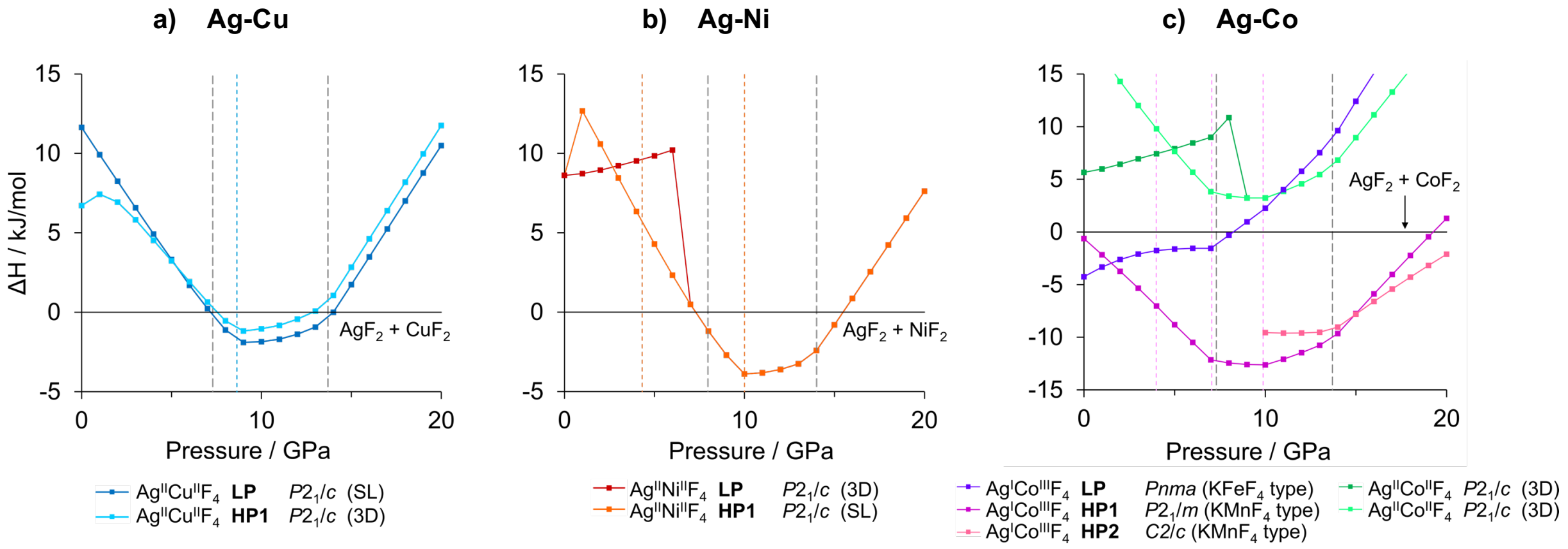

**Figure 3.** Calculated relative enthalpy of ternary silver fluorides $AgMF_4$ (M = Ni, Co, Cu) with respect to lowest-enthalpy polymorphic forms of substrates at each given pressure point. All the lowest enthalpy Ag(II)-M(II) fluorides comprise M-half-substituted $AgF_2$ type structures. Vertical dashed lines indicate phase transitions calculated for the $MF_2$ substrates. SL stands for the cation separated-layers polymorphs of $Ag^{II}M^{II}F_4$ while 3D denotes the spatially bonded.

## Evolution of stability with external pressure

The pressure evolution of enthalpy for $AgMF_4$ systems (M=Cu, Ni and Co) with respect to the sum of those for the binary substrates in their most stable polymorphs is presented in **Figure 3**. The phase-transition sequences in the considered stoichiometries were calculated with GGA+U assuming hydrostatic conditions.

$AgCuF_4$ HP1 polymorph (3D) is enthalpically favoured with respect to the substrates in the pressure range about 7-14 GPa with the minimum relative enthalpy at ca. 9 GPa (by –1.9 kJ/mol, **Figure 3a**). Similar pressure-induced stability is predicted for $AgNiF_4$ system HP1, with the separate layers (SL) structure gaining its stability under high-pressure (**Figure 3b**). The range of stability of the $AgNiF_4$ system is predicted to be about 7-15 GPa, with the maximum stabilization of –3.9 kJ/mol at 10 GPa relative to binary substrates. Due to small structural differences between both proposed structures of $AgNiF_4$ it may be expected, that if HP1 structure was synthesized under high-pressure it would decompress to their LP forms without large internal strain. In fact, during standard geometry optimization a smooth transition between both HP1 (SL) and LP (3D) polymorphs occurred (*i.e.* optimization of HP polymorph at 0 GPa gave LP structure and also optimization of LP polymorph at 10 GPa led to HP structure).

As mentioned before, the $Ag^{I}Co^{III}F_4$ system is stable with respect to divalent binary fluorides even at ambient pressure. Apparently, it turns out that $Ag^{I}Co^{III}F_4$ stability would increase almost threefold with respect to substrates at 10 GPa, up to –12.6 kJ/ mol, due to phase transition into monoclinic distorted $KMnF_4$ type polymorph HP1. Another, yet also monoclinic distorted $KMnF_4$ type polymorph HP2 (*C*2/*c*) with CN=10 for all Ag(I), was found to predominate the phase diagram of $Ag^{I}Co^{III}F_4$ system from 15 GPa for up to 22 GPa (**Figure 3c**, with linear extrapolation used to estimate transition pressure above 20 GPa). Interestingly, hypothetical $Ag^{II}Co^{II}F_4$ adopting $AgF_2$ type structures, *i.e.* similar to those of $AgCuF_4$ and $AgNiF_4$, would not be stabilized under high pressure (**Figure 3c** green curves).

Additional insight into pressure-induced effects on enthalpy of formation come from pV term analysis (**SI**, **Figure S3**). Both for relatively low pressure (< 3 GPa) and high pressure range (> 9 GPa) the pV factor works against formation of ternary fluoride phases. Simultaneously, ternary fluoride phases benefit from the pV term with respect to the substrates in the range 3–9 GPa; at higher pressures binary substrates experience volume-reducing phase transitions which reduces the said advantage. The crucial is the 2nd phase transition of $AgF_2$ at ~14 GPa, which reduces the pV term of substrates and gradually destabilizes thermodynamically all proposed ternary fluorides at pressures approaching 20

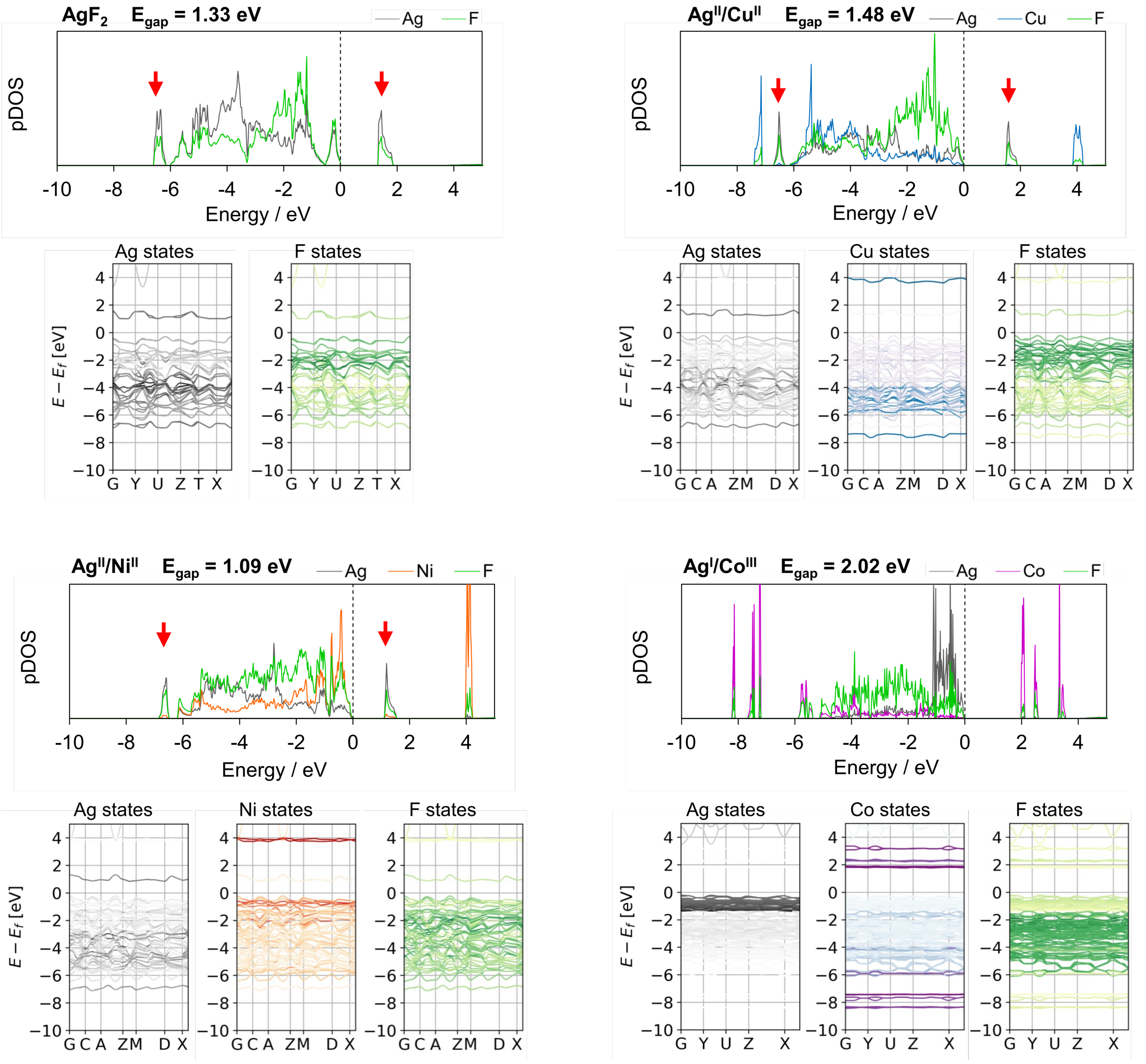

**Figure 4. Top:** Orbital projected electronic density of states (pDOS) of the parent $AgF_2$ and proposed $AgCuF_4$, $AgNiF_4$, $AgCoF_4$ compounds. Red arrows indicate lower and upper Hubbard (LHB and UHB) bands of Ag. **Bottom:** Band structure plots inside first Brillouin zone of the unit cells, where color intensity indicates respective element contribution (0-100%).

GPa. These factors result in re-entrant instability of ternary phases above certain characteristic pressure in each case.

## Electronic properties

Electronic density of states (DOS) and band structures for the title compounds and $AgF_2$ reference at ambient pressure are shown in **Figure 4**. Results of electronic structure calculations for the parent $AgF_2$ agree with previously published DFT+U results: it shows a narrow band gap (1.33 eV) with strongly mixture of Ag 4d and F 2p orbitals around the Fermi level[1,18]. Overall, $AgF_2$ is a charge-transfer (CT) insulator according to Zaanen-Sawatzky-Allen classification scheme.[34]

**Table 1.** Summary of the lowest-enthalpy structures predicted for $AgCoF_4$, $AgNiF_4$, $AgCuF_4$ and parent $AgF_2$ at ambient pressure, 10 GPa and 20 GPa. Calculated energies or enthalpies of formation, electronic band gaps and lengths of the three shortest Ag-F bonds are provided. The bond lengths describe $AgF_6$ octahedra distortion except in the case of $AgCoF_4$ structures. For the relative volume relation, see **Figure S3**.

| Structure at 0 GPa | | | $\Delta E_f$ / kJ/mol | $E_{gap}$ / eV | bond length / Å | | |
|---|---|---|---|---|---|---|---|
| System | Structure type | Symmetry | | | Ag-F$_1$ | Ag-F$_2$ | Ag-F$_3$ |
| $AgF_2$ | LP $AgF_2$ | *Pbca* | - | 1.334 | 2.072 | 2.075 | 2.560 |
| $AgCuF_4$ | LP $AgF_2$ type (SL) | *P*2$_1$/*c* | 6.7 | 1.484 | 2.049 | 2.071 | 2.598 |
| $AgNiF_4$ | LP $AgF_2$ type (3D) | *P*2$_1$/*c* | 8.6 | 0.902 | 2.025 | 2.260 | 2.332 |
| $AgCoF_4$ | LP $KFeF_4$ type | *Pnma* | -4.2 | 2.020 | 2.411 | 2.482 | 2.504 |

| Structure at 10 GPa | | | $\Delta H_f$ / kJ/mol | $E_{gap}$ / eV | bond length / Å | | |
|---|---|---|---|---|---|---|---|
| System | Structure type | Symmetry | | | Ag-F$_1$ | Ag-F$_2$ | Ag-F$_3$ |
| $AgF_2$ | HP1 $AgF_2$ | *Pca*2$_1$ | - | 1.311 | 2.056 | 2.061 | 2.369 |
| $AgCuF_4$ | HP1 $AgF_2$ type (3D) | *P*2$_1$/*c* | -1.9 | 1.512 | 2.061 | 2.105 | 2.604 |
| $AgNiF_4$ | HP1 $AgF_2$ type (SL) | *P*2$_1$/*c* | -3.9 | 0.618 | 2.021 | 2.112 | 2.433 |
| $AgCoF_4$ | HP1 $KMnF_4$ type dist. | *P*2$_1$/m | -12.6 | 1.786 | 2.313 | 2.379 | 2.411 |

| Structure at 20 GPa | | | $\Delta H_f$ / kJ/mol | $E_{gap}$ / eV | bond length / Å | | |
|---|---|---|---|---|---|---|---|
| System | Structure type | Symmetry | | | Ag-F$_1$ | Ag-F$_2$ | Ag-F$_3$ |
| $AgF_2$ | HP2 $AgF_2$ | *Pbcn* | - | 1.288 | 2.048 | 2.063 | 2.643 |
| $AgCuF_4$ | HP1 $AgF_2$ type (3D) | *P*2$_1$/*c* | 10.5 | 1.493 | 2.002 | 2.075 | 2.303 |
| $AgNiF_4$ | HP1 $AgF_2$ type (SL) | *P*2$_1$/*c* | 7.6 | 0.629 | 1.992 | 2.080 | 2.351 |
| $AgCoF_4$ | HP2 $KMnF_4$ type dist. | *C*2/*c* | -2.1 | 2.009 | 2.338 | 2.376 | 2.401 |

Analysis of $AgCuF_4$ electronic DOS at 0 GPa shows that Cu states are not admixing significantly to the states around Fermi level but they form a pair of Hubbard bands which are well-split from those characteristic of Ag. $AgCuF_4$ has a slightly broader band gap than parent $AgF_2$ (1.48 eV *vs.* 1.33 eV), while preserving its insulator character associated with a ligand to metal charge transfer (LMCT). Increasing the pressure up to 10 GPa and transition to the HP polymorph keeps the $AgCuF_4$ band gap nearly constant (**Figure S4**, at about 1.51 eV), as typical for the compounds of $d^9$ cations.

Calculations of electronic DOS for $AgNiF_4$ yield noticeably narrower band gap than in parent $AgF_2$ (1.09 eV). Interestingly, in this case valence states of all three constituting elements mix firmly within the valence band, but the Ag(II) contribution is now smaller than for pristine $AgF_2$. The valence band region is composed predominantly from F and Ni(II) states, while the conduction band preserves the character of Ag UHB band level. Under elevated pressure of 10 GPa the band gap of $AgNiF_4$ narrows down to 0.62 eV, which is less than half of parent $AgF_2$ at the same pressure (*cf.* HP evolution of electronic band gaps of ternary fluorides and $AgF_2$ in **SI**, **Figure S7**). These features suggest the inter-valence charge transfer (IVCT) character of $AgNiF_4$ and a certain propensity towards redox process yielding higher oxidation states of Ni, with a concomitant reduction of Ag(II) to Ag(I).

In the case of $AgCoF_4$ the band gap at ambient pressure is 2.02 eV thus much larger than those of pristine binary fluorides (**Figure 4**). The top of valence band mainly consists of filled d states of Ag(I) while the bottom of conduction band corresponds to the UHB of Co(III). Due to intrinsic redox reaction character, the band gap of $Ag^{I}Co^{III}F_4$ has an intervalence charge transfer (IVCT) character, with Ag(I) serving as an electron donor, and Co(III) as an acceptor. At 10 GPa band gap of $Ag^{I}Co^{III}F_4$ decreases noticeably down to 1.79 eV, but it is still about 0.5 eV larger than the corresponding gap for $AgF_2$ HP1 type.

Clearly, electronic structure of $AgMF_4$ compounds in the series M=Cu, Ni, Co, changes quite predictably with the redox properties of M(II) cations. For M=Cu, with Cu(III) oxidation state experimentally accessible with the greatest difficulty, $AgCuF_4$ preserves Ag(II) and Cu(II) oxidation states and a LMCT insulator character. For M=Ni, where Ni(III) oxidation state forms easier, electronic DOS of $AgNiF_4$ reveals a moderate ease of IVCT between Ni(II) donor and Ag(II) acceptor, with the corresponding valence and conduction bands separated by a mere ~1.1 eV. Finally, for M=Co, where Co(III) is most accessible, genuine redox reaction is seen which leads to "inverse" IVCT character – now Ag(I) serves as an electron donor, while Co(III) as an acceptor.

## 4. Conclusions

The silver(II) fluoride system is theoretically predicted to resist forming ternary fluorides with Cu, and Ni fluorides at ambient pressure conditions. The energy of formation of $Ag^{II}Cu^{II}F_4$ and $Ag^{II}Ni^{II}F_4$ are positive (+6.7 kJ/mol and +8.6 kJ/mol respectively), thus they could be only metastable if prepared. At ambient pressure, $Ag^{II}Cu^{II}F_4$ shows layered structure with buckled [$AgF_2$] layers, similar as in parent $AgF_2$. Also, the lowest-energy structure of $Ag^{II}Ni^{II}F_4$ stems from $AgF_2$ structure, but it forms 3D polymorph showing strong Ag-Ni bonds where Ag(II) acts like connector between $[NiF_{4/2}F_{2/1}]^{2-}$ layers. Both $Ag^{II}Cu^{II}F_4$ and $Ag^{II}Ni^{II}F_4$ are predicted to be semiconductors with predominant LMCT and IVCT character, respectively. $AgCuF_4$ system has a larger calculated band gap than that for pristine $AgF_2$ at ambient pressure (for $AgF_2$ 1.33 eV and 1.48 eV for $AgCuF_4$ LP), while $AgNiF_4$ system features a significantly narrower band gap (1.09 eV for $AgNiF_4$ LP).

The cobalt-silver fluoride system is quite different from the other two due to intrinsic redox reaction resulting in spontaneous formation of $Ag^{I}Co^{III}F_4$ in the $KFeF_4$ type. The ambient-pressure $AgCoF_4$ is computed to be a *ca.* 2.02 eV band gap semiconductor with IVCT character of the gap, featuring antiferromagnetic sheets of $[CoF_{4/2}F_{2/1}]^{-}$ stoichiometry, stable by –4.2 kJ/mol with respect to binary fluorides.

Application of hydrostatic pressure in our calculations shows that there is a pressure range where formation of ternary silver fluorides $Ag^{II}M^{II}F_4$ is favoured. The range of stability is about 7-15 GPa for $AgCuF_4$, 8-15 GPa for $AgNiF_4$, and 0-22 GPa for $AgCoF_4$. Nickel system is noteworthy, because at 10 GPa it should form the separate layers polymorph with quite narrow band gap (0.62 eV, less than half of that for $AgF_2$) and simultaneously slightly negative enthalpy of formation –3.9 (kJ/mol) with respect to substrates. Finally, for $AgCoF_4$ the ambient pressure $KFeF_4$-LT type is calculated to yield to distorted monoclinic $KMnF_4$ type structure in the pressure range of 1.5-19 GPa.

**Acknowledgments**: This work was supported by Polish National Science Center (NCN) within Beethoven project (2016/23/G/ST5/04320). The research was carried out using supercomputers of Interdisciplinary Centre for Mathematical and Computational Modelling (ICM), University of Warsaw, under grant number GA76-19. Dr Derzsi acknowledges the ERDF, R&I Operational Program (ITMS2014+: 313011W085), Scientific Grant Agency of the Slovak Republic grant (VG 1/0223/19) and the Slovak Research and Development Agency grant (APVV-18-0168).

# Supporting information

## Theoretical study of ternary silver fluorides $AgMF_4$ (M = Co, Ni, Cu) formation at pressures up to 20 GPa

M.A. Domański[1]*, M. Derzsi[1,2] and W. Grochala[1]*

**Contents**

### S1. Estimated phase-transition pressures for the substrates

The calculations were performed for the pressures up to 20 GPa, thus we considered the known phase transitions (PTs) for $AgF_2$[19] (sequence $Pbca$ → $Pca2_1$ (~8 GPa) → $Pbcn$ (~14 GPa)), AgF[35] (NaCl-type → CsCl-type (1-3 GPa)), $CoF_2$[16,21] (rutile → $Pnnm$ (~4 GPa) → $Pbca/Pa\overline{3}$ (~8 GPa) → $I4/mmm$ (~12 GPa)), $NiF_2$[21] (rutile → $Pnnm$ (~4 GPa) → $I4/mmm$ (~10 GPa)) and $CuF_2$[20] ($P2_1/c$ → $Pbca$ (~9 GPa)).

Our results are presented on the **Figure S1** and in the **Table S1**. All of the abovementioned PTs were confirmed using chosen theoretical approach, resulting in PT pressure differences smaller than 1.0 GPa (the only exception is nearly 2 GPa shift for 3rd PT of $CoF_2$). We indicate also that so called in literature "distorted fluorite" HP polymorph of $NiF_2$ (assumed $I4/mmm$) would have too high enthalpy. Instead, we propose formation of orthorhombic distorted $PdF_2$ type structure (*Pbca* symmetry), which is consistent with $CoF_2$ and $CuF_2$ sequences of pressure-induced PTs.

## S2. Supplementary Figures

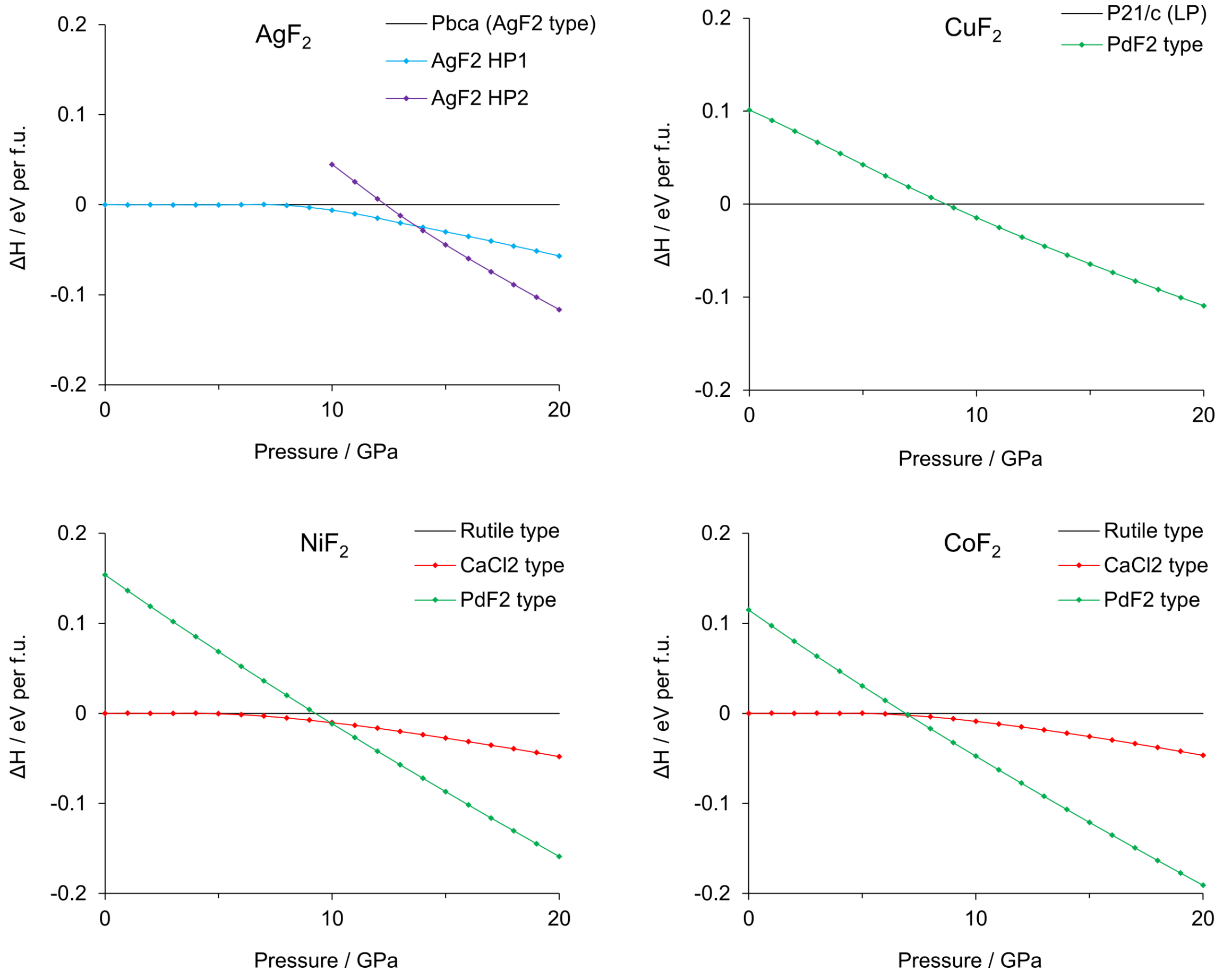

**Figure S1.** Enthalpy versus pressure diagrams for parent binary fluorides.

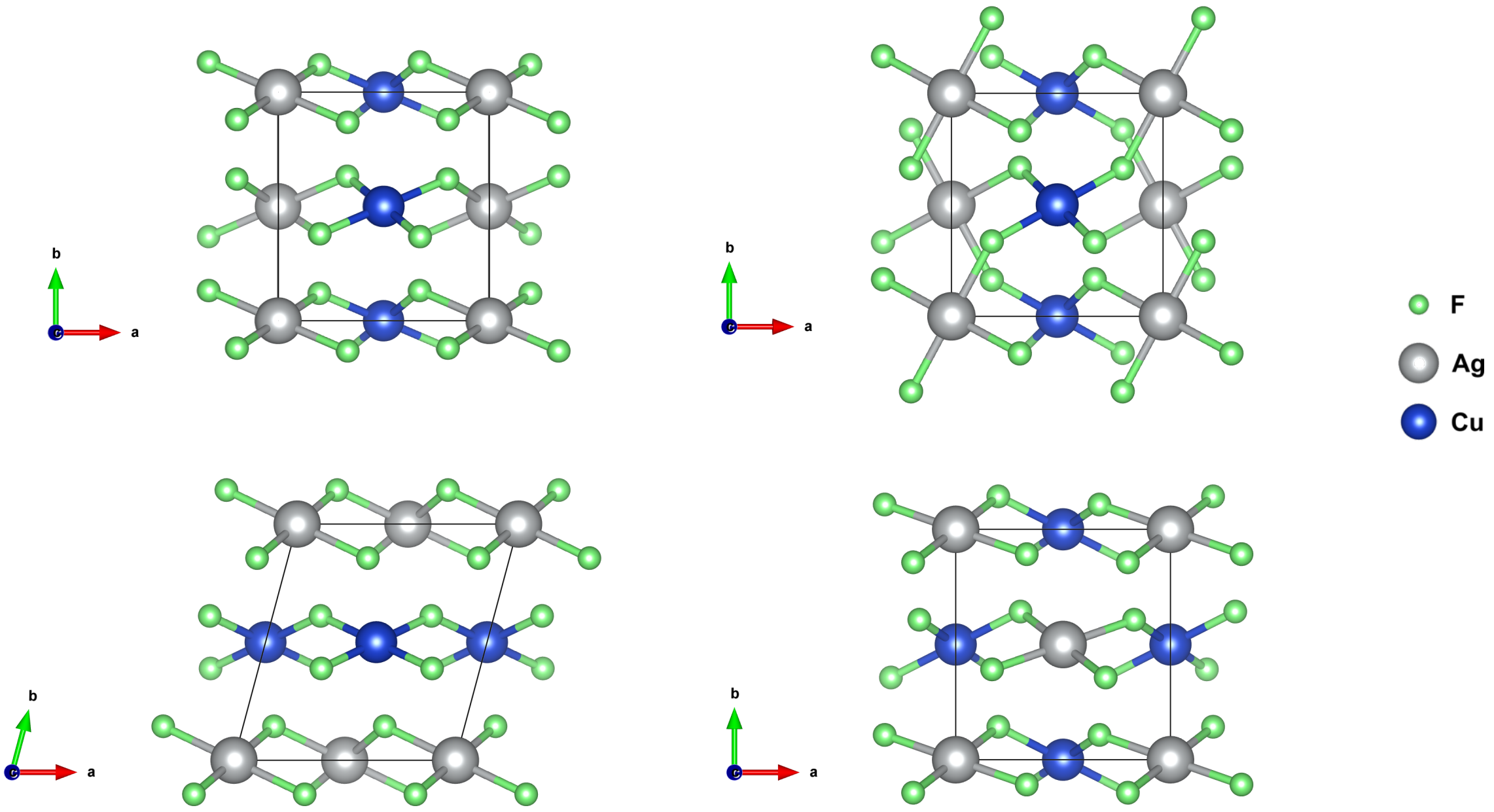

**Figure S2.** Structures of $AgCuF_4$ originating from various substitutions within $AgF_2$ type structure enabled by its orthorhombic symmetry. Starting from top-left corner: $\perp\vec{a}$, $\perp\vec{a}$ (3D, HP1 polytype), $\perp\vec{b}$ (SL, LP polytype) and $\perp\vec{c}$.

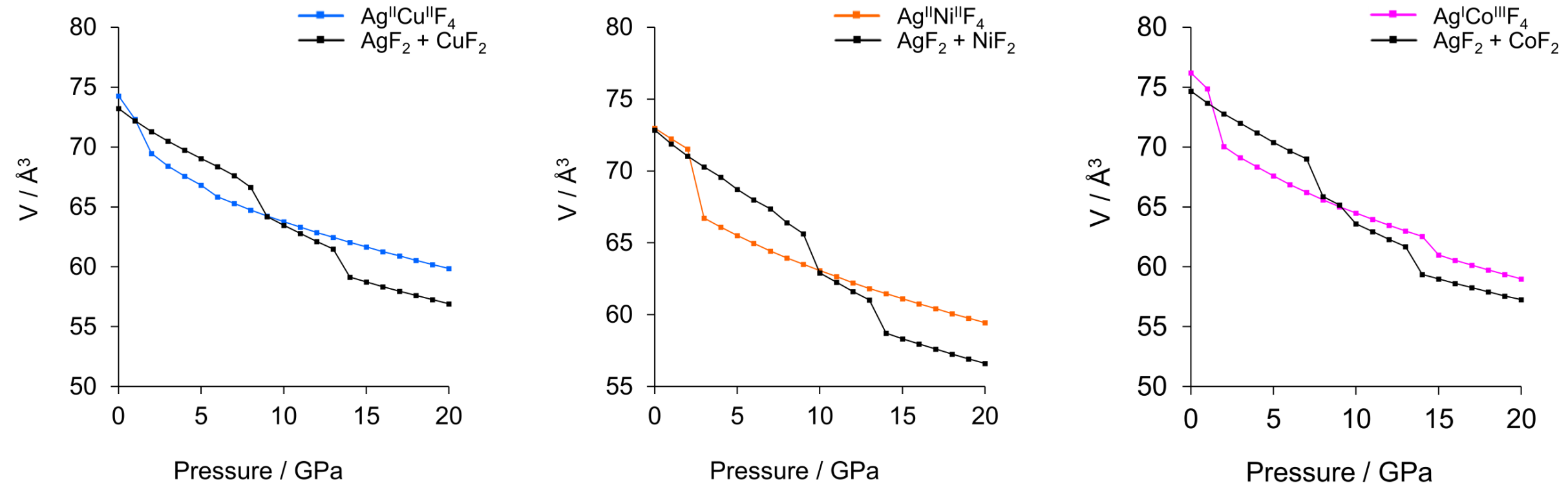

**Figure S3.** Pressure versus volume plots for the lowest-enthalpy polymorphs of $AgCuF_4$, $AgNiF_4$ and $AgCoF_4$ with respect to the considered substrates.

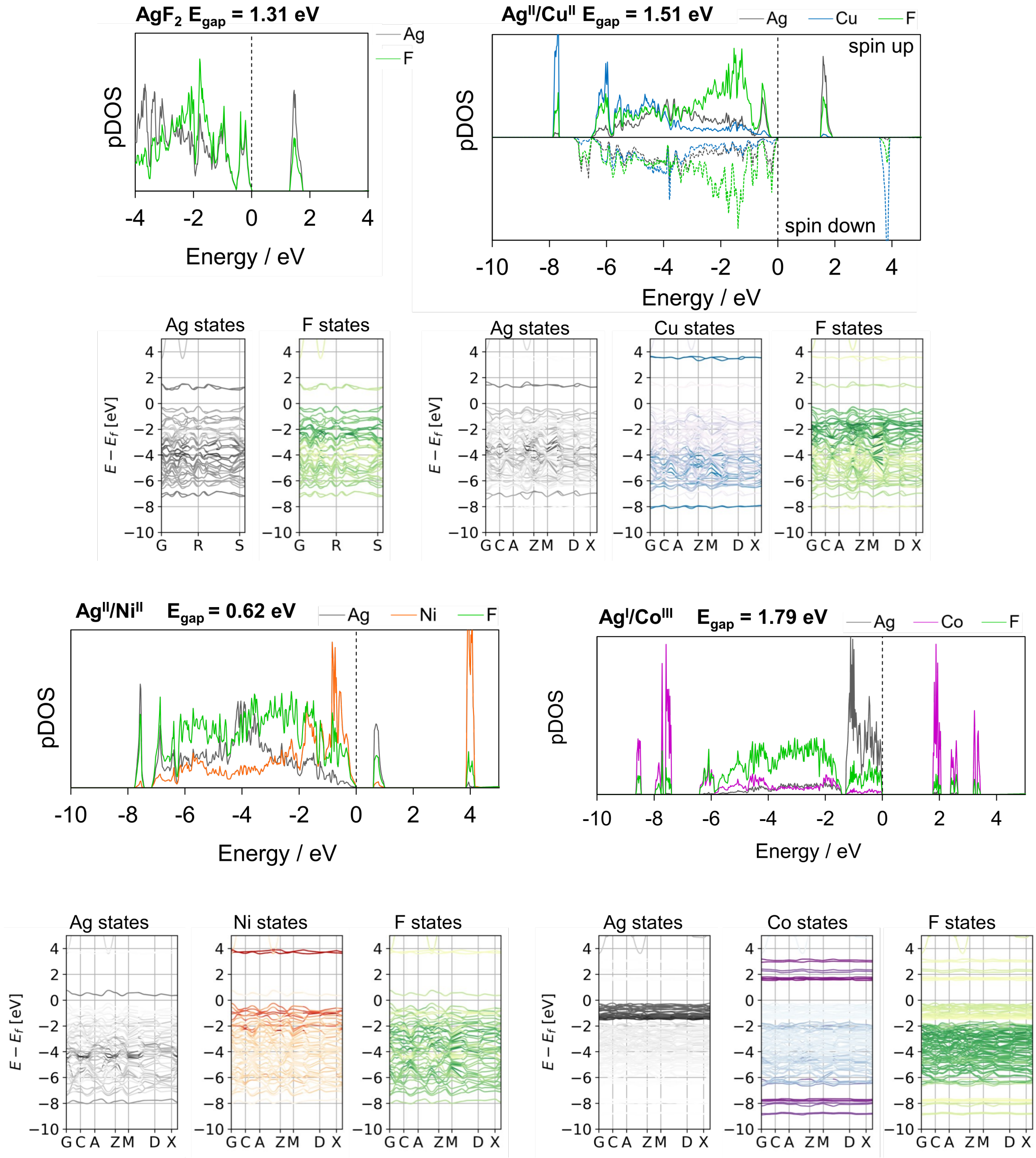

**Figure S4**. Electronic density of states and band structures of $AgF_2$, $AgCuF_4$, $AgNiF_4$ and $AgCoF_4$ at 10 GPa.

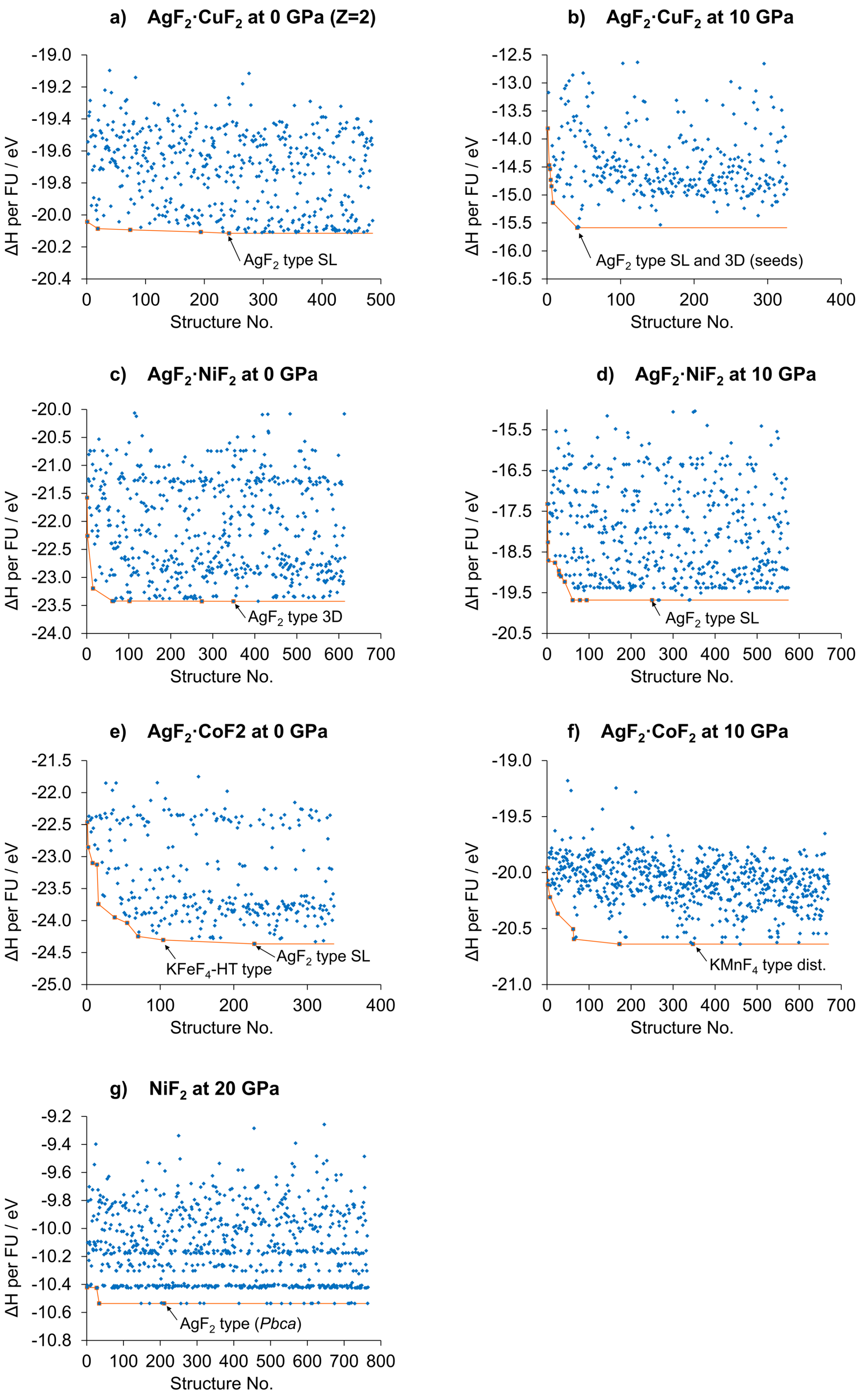

**Figure S5.** Performed XtalOpt quests for the lowest-enthalpy structures for the $AgCuF_4$, $AgNiF_4$, $AgCoF_4$ and $NiF_2$ stoichiometries for Z=4 (with the one exception marked for $AgF_2{\cdot}CuF_2$ at 0 GPa). All structures energies are marked with blue diamonds, while the lowest-enthalpy structures are marked with orange squares and are connected with lines.

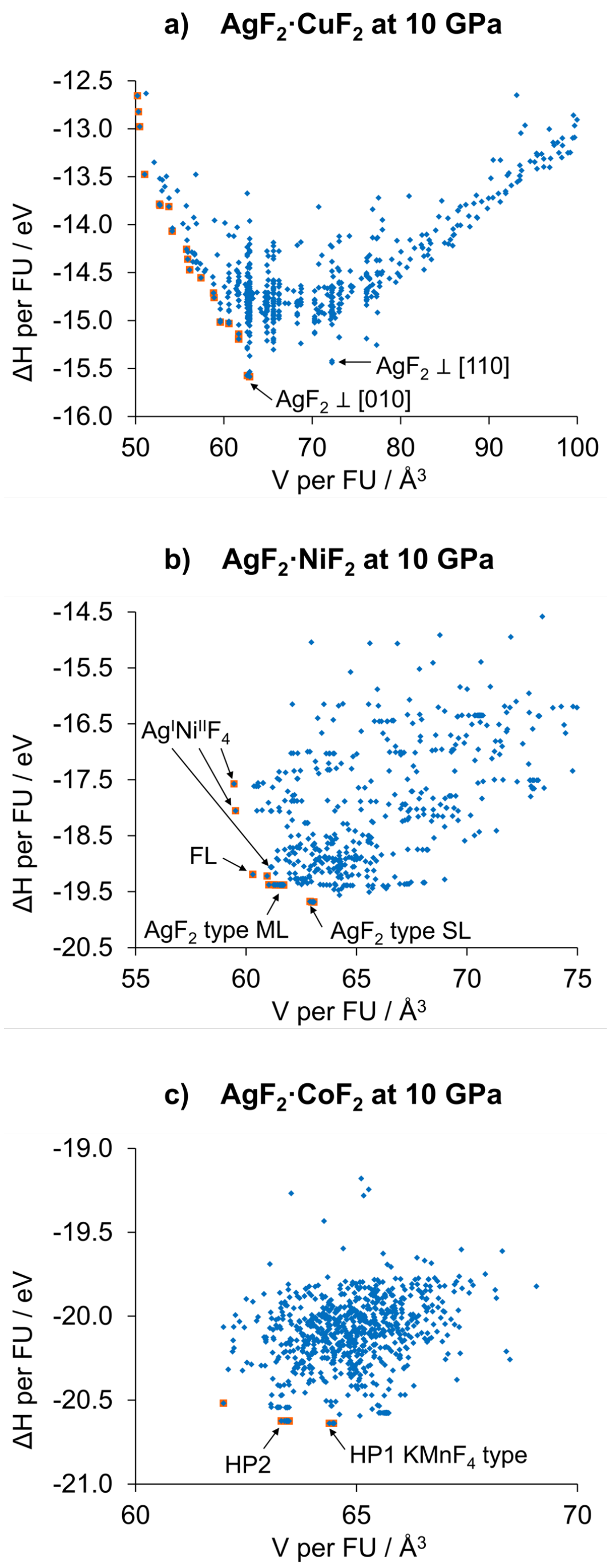

**Figure S6.** Performed XtalOpt quests for the lowest-enthalpy structures for the $AgCuF_4$, $AgNiF_4$ and $AgCoF_4$ stoichiometries for Z=4 at 10 GPa. All structures energies are marked with blue diamonds, while the lowest enthalpy & volume structures are marked with orange squares. On the plot **a)** given crystal directions mark the substitution plane in particular structures. On the plot **b)** FL denotes flat layers (mixed Ag-Ni layers) polymorph, ML buckled layers (mixed Ag-Ni layers) polymorph, while SL separate layers polymorph (*i.e.* element separated).

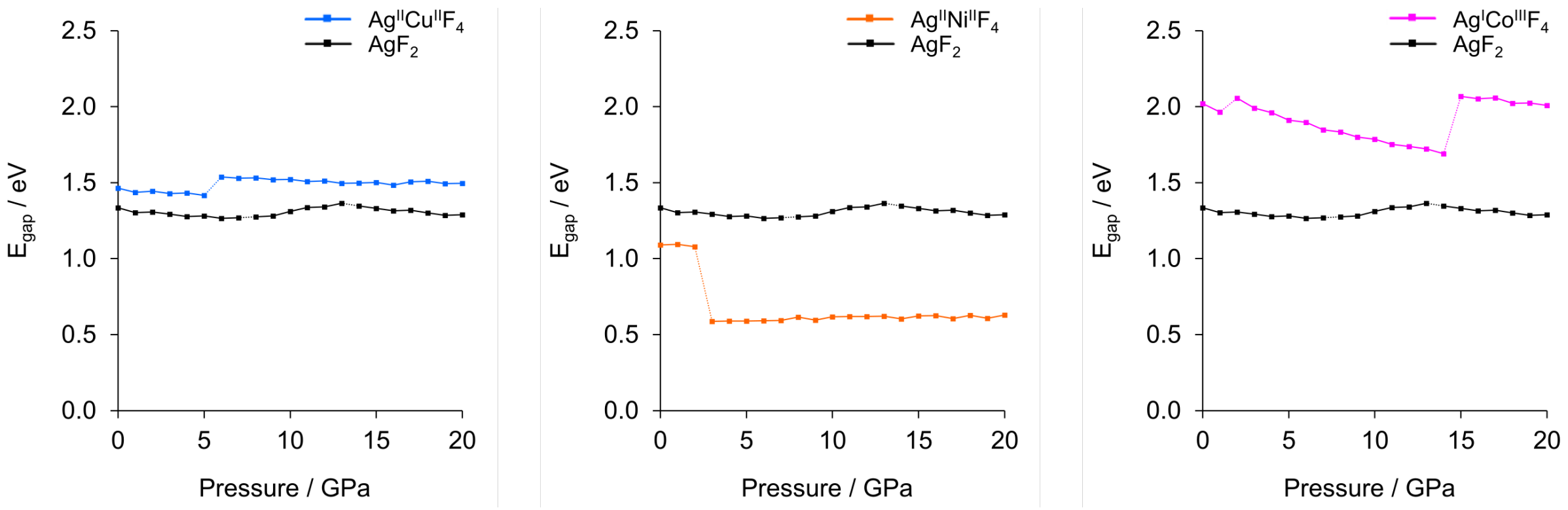

**Figure S7.** Electronic band gaps dependency on pressure increase for the lowest-enthalpy structures of $AgCuF_4$, $AgNiF_4$ and $AgCoF_4$ (shown together with $AgF_2$). Dotted lines connect structures separated by corresponding phase transitions. Detailed data is provided in the **Table S3**.

## S3. Supplementary Tables

**Table S1.** Phase transition pressures obtained from GGA+U calculations for parent binary fluorides compared with experimental data.

Substrates phase transitions < 20 GPa

| $AgF_2$ | | |
|---|---|---|
| PT | Calc. | Exp. |
| I | 7.3 | 7.8 [19] |
| II | 13.7 | 14.2 [19] |

| $CoF_2$ | | |
|---|---|---|
| PT | Calc. | Exp. |
| I | 4.0 | 3.6 [16] |
| II | 7.0 | 8 [16] |
| III | 9.9 | 12 [16] |

| $NiF_2$ | | |
|---|---|---|
| PT | Calc. | Exp. |
| I | 4.4 | 4-5 [21] |
| II | 9.9 | 10 [21] |

| $CuF_2$ | | |
|---|---|---|
| PT | Calc. | Exp. |
| I | 8.6 | 9 [20] |

*all data given in GPa

**Table S2.** Comparison of GGA+U and hybrid HSE06 energies of formation and volume for presented ternary fluorides and parent phases of binary fluorides.

| **Structure at 0 GPa** | | | **GGA+U** | | | **HSE06** | | |
|---|---|---|---|---|---|---|---|---|
| **System** | **Structure type** | **Symmetry** | **ΔE$_f$ / kJ/mol** | **V / Å$^3$** | $V_{prod}/V_{substr}$ | **ΔE$_f$ / kJ/mol** | **V / Å$^3$** | $V_{prod}/V_{substr}$ |
| $AgCoF_4$ | $KFeF_4$ type | *Pnma* | -4.2 | 76.19 | 102.0% | -24.8<br>*-37.1 | 77.32 | 103.0% |
| $AgNiF_4$ | $AgF_2$ type (3D) | *P*2$_1$/*c* | 8.6 | 72.98 | 100.2% | 12.0 | 73.38 | 99.8% |
| $AgCuF_4$ | $AgF_2$ type (SL) | *P*2$_1$/*c* | 6.7 | 74.25 | 101.4% | 6.1 | 75.78 | 101.3% |
| $AgF_2$ | $AgF_2$ type | *Pbca* | - | 40.48 | - | - | 41.06 | - |
| $CuF_2$ | $CuF_2$ type | *P*2$_1$/*c* | - | 32.74 | - | | 33.77 | |
| $NiF_2$ | Rutile type | *P*4/*mmm* | - | 32.37 | - | | 32.46 | |
| $CoF_2$ | Rutile type | *P*4/*mmm* | - | 34.19 | - | | 34.03 | |

*with respect to the $AgF + CoF_3$ reaction

| **Structure at 10 GPa** | | | **GGA+U** | | | **HSE06** | | |
|---|---|---|---|---|---|---|---|---|
| **System** | **Structure type** | **Symmetry** | **ΔH$_f$ / kJ/mol** | **V / Å$^3$** | $V_{prod}/V_{substr}$ | **ΔH$_f$ / kJ/mol** | **V / Å$^3$** | $V_{prod}/V_{substr}$ |
| $AgCoF_4$ | $KMnF_4$ type dist. | *P*2$_1$/*m* | -12.6 | 64.49 | 101.4% | -26.4 | 65.39 | 103.0% |
| $AgNiF_4$ | $AgF_2$ type (SL) | *P*2$_1$/*c* | -3.9 | 63.07 | 100.2% | -2.1 | 63.57 | 100.2% |
| $AgCuF_4$ | $AgF_2$ type (3D) | *P*2$_1$/*c* | -1.9 | 63.78 | 100.5% | -3.5 | 64.94 | 100.4% |
| $AgF_2$ | $AgF_2$ HP1 type | *Pca*2$_1$ | - | 35.19 | - | - | 35.62 | - |
| $CuF_2$ | $AgF_2$ type | *Pbca* | - | 28.28 | - | - | 29.06 | - |
| $NiF_2$ | $AgF_2$ type | *Pbca* | - | 27.73 | - | - | 27.83 | - |
| $CoF_2$ | $AgF_2$ type | *Pbca* | - | 28.40 | - | - | 28.16 | - |

**Table S3.** Electronic band gaps dependency on pressure increase for the lowest-enthalpy structures of $AgCuF_4$, $AgNiF_4$, $AgCoF_4$ and $AgF_2$.

| p / GPa | $AgF_2$ | | $AgCuF_4$ | | $AgNiF_4$ | | $AgCoF_4$ | |
|---|---|---|---|---|---|---|---|---|
| | $E_{gap}$ / eV | Structure | $E_{gap}$ / eV | Structure | $E_{gap}$ / eV | Structure | $E_{gap}$ / eV | Structure |
| 0 | 1.334 | LP | 1.463 | LP (SL) | 1.091 | LP (3D) | 2.020 | LP |
| 1 | 1.302 | LP | 1.437 | LP (SL) | 1.094 | LP (3D) | 1.965 | LP |
| 2 | 1.307 | LP | 1.444 | LP (SL) | 1.079 | LP (3D) | 2.057 | HP1 |
| 3 | 1.293 | LP | 1.429 | LP (SL) | 0.587 | HP1 (SL) | 1.991 | HP1 |
| 4 | 1.277 | LP | 1.433 | LP (SL) | 0.589 | HP1 (SL) | 1.961 | HP1 |
| 5 | 1.281 | LP | 1.417 | LP (SL) | 0.590 | HP1 (SL) | 1.911 | HP1 |
| 6 | 1.266 | LP | 1.538 | HP1 (3D) | 0.591 | HP1 (SL) | 1.898 | HP1 |
| 7 | 1.269 | LP | 1.530 | HP1 (3D) | 0.593 | HP1 (SL) | 1.847 | HP1 |
| 8 | 1.275 | HP1 | 1.532 | HP1 (3D) | 0.615 | HP1 (SL) | 1.833 | HP1 |
| 9 | 1.282 | HP1 | 1.519 | HP1 (3D) | 0.596 | HP1 (SL) | 1.800 | HP1 |
| 10 | 1.311 | HP1 | 1.522 | HP1 (3D) | 0.618 | HP1 (SL) | 1.786 | HP1 |
| 11 | 1.337 | HP1 | 1.507 | HP1 (3D) | 0.620 | HP1 (SL) | 1.753 | HP1 |
| 12 | 1.341 | HP1 | 1.511 | HP1 (3D) | 0.620 | HP1 (SL) | 1.738 | HP1 |
| 13 | 1.364 | HP1 | 1.495 | HP1 (3D) | 0.622 | HP1 (SL) | 1.723 | HP1 |
| 14 | 1.347 | HP2 | 1.497 | HP1 (3D) | 0.603 | HP1 (SL) | 1.690 | HP1 |
| 15 | 1.330 | HP2 | 1.501 | HP1 (3D) | 0.624 | HP1 (SL) | 2.069 | HP2 |
| 16 | 1.314 | HP2 | 1.484 | HP1 (3D) | 0.625 | HP1 (SL) | 2.053 | HP2 |
| 17 | 1.318 | HP2 | 1.505 | HP1 (3D) | 0.606 | HP1 (SL) | 2.058 | HP2 |
| 18 | 1.301 | HP2 | 1.509 | HP1 (3D) | 0.627 | HP1 (SL) | 2.022 | HP2 |
| 19 | 1.285 | HP2 | 1.493 | HP1 (3D) | 0.608 | HP1 (SL) | 2.025 | HP2 |
| 20 | 1.288 | HP2 | 1.495 | HP1 (3D) | 0.629 | HP1 (SL) | 2.009 | HP2 |

## S4. Crystal structures in POSCAR format

```
AgCuF4 SL at 0 GPa (P2_1/c)
   1.00000000000000
     5.3901036487284317    0.7313930884644128    0.0000000000000000
     0.7999934734740585    5.8918165998868357    0.0000000000000000
     0.0000000000000000    0.0000000000000000    4.7638498089899999
   Ag    Cu    F
     2     2     8
Direct
 -0.0000000000000000 -0.0000000000000000 -0.0000000000000000
  0.5000000000000000 -0.0000000000000000  0.5000000000000000
 -0.0000000000000000  0.5000000000000000  0.5000000000000000
  0.5000000000000000  0.5000000000000000 -0.0000000000000000
  0.3610222200924554  0.8549158998614186  0.1646245680338823
  0.6389777799075446  0.1450840701385790  0.8353754559661197
  0.1389777799075446  0.1450840701385790  0.6646245910338806
  0.8610222630924554  0.8549158998614186  0.3353754559661197
  0.7865728171829347  0.3876784498046880  0.2457279423781989
  0.2134271828170653  0.6123215501953191  0.7542721046218013
  0.7134272258170653  0.6123215501953191  0.7457279423781991
  0.2865728171829346  0.3876784498046880  0.2542720816218031
```

```
AgCuF4 3D at 10 GPa (P2_1/c)
   1.00000000000000
     4.9274005811963182    0.0000000000000000   -0.0515194211623682
     0.0000000000000000    5.1535447840790782    0.0000000000000000
    -0.0654618047501239    0.0000000000000000    5.0237402810958489
   Ag    Cu    F
     2     2     8
Direct
  0.0000000000000000 -0.0000000000000000 -0.0000000000000000
  0.0000000000000000  0.5000000000000000  0.5000000000000000
  0.5000000000000000  0.5000000000000000 -0.0000000000000000
  0.5000000000000000 -0.0000000000000000  0.5000000000000000
  0.3389463792717425  0.8347599318493771  0.1960332912334153
  0.6610536207282576  0.1652400381506133  0.8039667327665866
  0.1779727376733418  0.1555907884328188  0.5930914897534010
  0.8220273053266582  0.8444091815671787  0.4069085572465994
  0.6610536207282576  0.3347599718493946  0.3039667327665866
  0.3389463792717425  0.6652400281506196  0.6960333142334137
  0.8220273053266582  0.6555907784328180  0.9069085572465994
  0.1779727376733418  0.3444092215671749  0.0930914667534026
```

```
AgNiF4 3D at 0 GPa (P2_1/c)
1.0
        5.6236634254         0.0000000000         0.0000000000
        0.0000000000         4.8052668571         0.0000000000
       -1.6601559382         0.0000000000         5.4010045697
   Ag   Ni    F
    2    2    8
Direct
     0.000000000         0.000000000         0.000000000
     0.000000010         0.500000000         0.500000021
     0.500000000         0.500000000         0.000000000
     0.500000021        -0.000000000         0.500000021
     0.631682708         0.291337303         0.316754088
     0.368317340         0.708662772         0.683245975
     0.827889647         0.830939754         0.599213142
     0.172110334         0.169060234         0.400786943
     0.368316524         0.791337526         0.183244983
     0.631683475         0.208662449         0.816755058
     0.172111236         0.330939109         0.900784888
     0.827888830         0.669060842         0.099215164
```

```
AgNiF4 SL at 10 GPa (P2_1/c)
   1.00000000000000
     5.0886782073707630    0.0000000000000000    0.1368241482182605
     0.0000000000000000    4.8499977824595426    0.0000000000000000
    -0.8194022892055097    0.0000000000000000    5.0886650244517631
   Ag   Ni   F
     2     2     8
Direct
  0.5000000100000008  0.0000000000000000  0.5000000000000000
  0.5000000000000000  0.5000000000000000  0.0000000000000000
 -0.0000000000000000  0.0000000000000000 -0.0000000000000000
  0.9999999990000035  0.5000000000000000  0.5000000000000000
  0.8690903136360084  0.1781934389142069  0.2948350645973958
  0.1309096353639909  0.8218065730857941  0.7051649114026022
  0.1309096403639878  0.6781934269142059  0.2051649344026005
  0.8690903276360096  0.3218065730857941  0.7948350875973942
  0.3441785184272516  0.3509922888094633  0.6260187667631506
  0.6558214965727461  0.6490076871905347  0.3739812102368510
  0.6558215105727473  0.8509922638094648  0.8739812322368529
  0.3441785114272546  0.1490077231905305  0.1260187787631515
```

```
AgCoF4 KFeF4-LT-type at 0 GPa (Pnma)
   1.00000000000000
     11.0229778442210424    0.0000000000000000    0.0000000000000000
      0.0000000000000000    7.3311291705040329    0.0000000000000000
      0.0000000000000000   -0.0000000000000000    7.5425856227456265
   Ag   Co   F
     8     8    32
Direct
  0.9683188982466118  0.2500000000000000  0.1255183651661389
  0.0316810847533832  0.7500000160000013  0.8744816728338642
  0.9837239260802026  0.2500000000000000  0.6228451327222728
  0.0162760599197891  0.7500000160000013  0.3771549282777215
  0.5316811017533882  0.7500000160000013  0.6255183571661382
  0.4683188982466118  0.2500000000000000  0.3744816428338617
  0.5162760739197974  0.7500000160000013  0.1228450867222762
  0.4837239260802098  0.2500000000000000  0.8771548982777262
  0.7467105127796945  0.4990345616993822  0.3738868254660963
  0.7532894872203055  0.9990345616993822  0.8738868864660977
  0.7532894872203055  0.5009654383006178  0.8738868864660977
  0.7467105127796945  0.0009654253006132  0.3738868254660963
  0.2532895072203000  0.9990345616993822  0.6261131435339048
  0.2467105127797016  0.4990345616993822  0.1261131665339031
  0.2467105127797016  0.0009654253006132  0.1261131665339031
  0.2532895072203000  0.5009654383006178  0.6261131435339048
  0.7539609126181328  0.0535814051346022  0.1238451037124495
  0.2460390493818711  0.9464186048653986  0.8761549272875495
  0.7460390873818672  0.9464186048653986  0.6238451037124495
  0.2539608926181312  0.0535814051346022  0.3761548962875505
  0.2460390493818711  0.5535814261346075  0.8761549272875495
  0.7539609126181328  0.4464185738653925  0.1238451037124495
  0.2539608926181312  0.4464185738653925  0.3761548962875505
  0.7460390873818672  0.5535814261346075  0.6238451037124495
  0.1963206150647456  0.2500000000000000  0.0795629127220840
  0.8036793849352544  0.7500000160000013  0.9204370722779112
  0.3036793649352528  0.7500000160000013  0.5795628977220864
  0.6963206150647456  0.2500000000000000  0.4204370722779112
  0.2938130443724235  0.7500000160000013  0.1739096856442099
  0.7061869556275694  0.2500000000000000  0.8260903443557925
  0.2061869556275765  0.2500000000000000  0.6739096856442099
  0.7938130443724306  0.7500000160000013  0.3260903143557901
  0.4070914288000720  0.0605024491404656  0.1132071308650861
  0.5929085711999279  0.9394975318595364  0.8867929071349100
  0.0929085711999280  0.9394975318595364  0.6132071228650925
  0.9070913908000761  0.0605024491404656  0.3867928771349146
  0.5929085711999279  0.5605025001404662  0.8867929071349100
  0.4070914288000720  0.4394975628595353  0.1132071308650861
  0.9070913908000761  0.4394975628595353  0.3867928771349146
  0.0929085711999280  0.5605025001404662  0.6132071228650925
  0.0859978345849887  0.9426344231570277  0.1375159744150148
  0.9140022044150110  0.0573655528429703  0.8624839795849886
  0.4140021654150113  0.0573655528429703  0.6375159894150124
  0.5859977955849890  0.9426344231570277  0.3624840105849875
  0.9140022044150110  0.4426343921570288  0.8624839795849886
  0.0859978345849887  0.5573655448429696  0.1375159744150148
```

```
  0.5859977955849890  0.5573655448429696  0.3624840105849875
  0.4140021654150113  0.4426343921570288  0.6375159894150124


AgCoF4 KMnF4-type monoclinic distorted at 10 GPa (P2_1/m)
   1.00000000000000
     5.1645581604201301   -0.0000000000000000    -0.0000000000000000
    -0.0000000000000000    7.1288208007760936   -0.0000000000000000
    -0.2681686183553522   -0.0000000000000000    7.0060921224724524
   Ag   Co   F
     4     4    16
Direct
  0.5012381913199817  0.7500003415348069  0.1765428341331357
  0.4987584773613909  0.2499994627704109  0.8234548225013257
  0.5381112012013892  0.7499996533152838  0.6908786788800290
  0.4618853107046126  0.2500005035353399  0.3091227344654247
 -0.0000080995564997 -0.0000008442355662  0.4999975127345845
 -0.0000136385151722  0.4999966396965198  0.0000037605234955
 -0.0000053607764640  0.5000003312156946  0.4999980065919969
 -0.0000158398237898  0.0000035478031278  0.0000043805271134
  0.8618168068943673  0.7499999387999645  0.9640632722255620
  0.1381785410167011  0.2499997827395836  0.0359365976662595
  0.0792269506274666  0.7500002627852215  0.5860037788829671
  0.9207721279647567  0.2499998008160619  0.4139949585530782
  0.8640631688274300  0.0550811181571014  0.7465654681665583
  0.3048455422020692  0.9313245384715036  0.9115906634229285
  0.3172011439669050  0.4282452760175758  0.5771973510137838
  0.8640608187847351  0.4449219082362519  0.7465646471760112
  0.6951527704141568  0.4313202880577797  0.0884028473736006
  0.1359367482347910  0.5550776190818745  0.2534349664641848
  0.1359326286795765  0.9449190249828796  0.2534343815857265
  0.6827992686805414  0.9282422440873904  0.4228086446263350
  0.6828011167966012  0.5717550420208358  0.4228025545828132
  0.3048465151447165  0.5686799473612165  0.9115951781725444
  0.3172010542296453  0.0717584667452880  0.5771919250978584
  0.6951523036200709  0.0686742860038426  0.0884073746326691
```

```
 AgCoF4 KMnF4-type monoclinic distorted at 20 GPa (C2/c)
 1.00000000000000
    11.0812628404247011    0.0000000000000000   -0.0059732377468867
     0.0000000000000000    4.4484920880475007    0.0000000000000000
    -1.5545900041669392    0.0000000000000000    4.7871892623391794
   Ag   Co   F
     4     4    16
Direct
 -0.0000000000000000  0.9854499414981792 -0.0000000000000000
  0.5000000000000000  0.4854499414981792 -0.0000000000000000
  0.0000000049999969  0.6930680766525872  0.5000000189999980
  0.5000000260000022  0.1930680386525841  0.5000000189999980
  0.7499826645943405  0.0891191538441723  0.2499874905508578
  0.2499826645943404  0.5891191278441701  0.2499874905508578
  0.7500173424056565  0.5891191278441701  0.7500125234491434
  0.2500173524056575  0.0891191538441723  0.7500125234491434
  0.9046789962279855  0.4494612861620155  0.7950579031525441
  0.4046789542279891  0.9494613381620199  0.7950579031525441
  0.9046364545787734  0.2288860483661350  0.2951229472303552
  0.4046364965787698  0.7288860743661372  0.2951229472303552
  0.0953635414212262  0.2288860483661350  0.7048770657696495
  0.5953635514212271  0.7288860743661372  0.7048770657696495
  0.0953210617720122  0.4494612861620155  0.2049421348474519
  0.5953210977720151  0.9494613381620199  0.2049421348474519
  0.6901823796910656  0.4408622789629895  0.3802029965247833
  0.1901823796910656  0.9408622789629895  0.3802029965247833
  0.6901711840959881  0.2374122220893062  0.8801913918524845
  0.1901711530959890  0.7374122220893062  0.8801913918524845
  0.8098288629040051  0.7374122220893062  0.1198086101475157
  0.3098288409040105  0.2374122220893062  0.1198086101475157
  0.8098176273089386  0.9408622789629895  0.6197970174752178
  0.3098176273089385  0.4408622789629895  0.6197970174752178
```

```
NiF2 orthorhombic PdF2-type at 20 GPa (Pbca)
   1.00000000000000
     4.7184679081789511   -0.0000000000000000    0.0000000000000000
     0.0000000000000000    4.7140006131837016    0.0000000000000000
     0.0000000000000000    0.0000000000000000    4.7425333960465874
   Ni   F
     4     8
Direct
 -0.0000000000000000 -0.0000000000000000 -0.0000000000000000
  0.5000000000000000 -0.0000000000000000  0.5000000000000000
  0.0000000000000000  0.5000000000000000  0.5000000000000000
  0.5000000000000000  0.5000000000000000 -0.0000000000000000
  0.3483211307683596  0.3482448345408198  0.3468214467687561
  0.6516788462316349  0.6517551424591748  0.6531785302312455
  0.1516788692316403  0.6517551424591748  0.8468214697687545
  0.8483211537683651  0.3482448345408198  0.1531785532312439
  0.6516788462316349  0.8482448575408252  0.1531785532312439
  0.3483211307683596  0.1517551654591802  0.8468214697687545
  0.8483211537683651  0.1517551654591802  0.6531785302312455
  0.1516788692316403  0.8482448575408252  0.3468214467687561
```